\documentclass [aps,prb,twocolumn,superscriptaddress]{revtex4-2}
\usepackage{graphicx}
\usepackage{amsmath}
\usepackage{braket}
\usepackage{xcolor}
\usepackage{bm}
\usepackage{ulem}
\usepackage{physics}
\usepackage{mathtools}
\usepackage[unicode]{hyperref}
\usepackage{enumerate}
\hypersetup{
	colorlinks   = true, %Color links instead of boxes
	urlcolor     = black, %Color for external hyperlinks
	linkcolor    = black, %Coloψηαr of internal links
	citecolor   = black, %Color of citations
}

\begin{document}
\title{
\textit{Ab initio} investigation of the topological Hall effect\\ caused by magnetic skyrmions in Pd/Fe/Ir(111)
}

\author{Adamantia Kosma}
\affiliation{Peter Gr\"unberg Institut (PGI-1), Forschungszentrum J\"ulich and JARA, 52425 J\"ulich, Germany}
\author{Philipp R\"u\ss mann}
\affiliation{Peter Gr\"unberg Institut (PGI-1), Forschungszentrum J\"ulich and JARA, 52425 J\"ulich, Germany}  
\affiliation{Institute for Theoretical Physics and Astrophysics,
University of W\"urzburg, 97074 W\"urzburg, Germany}
\author{Yuriy Mokrousov}
\affiliation{Peter Gr\"unberg Institut (PGI-1), Forschungszentrum J\"ulich and JARA, 52425 J\"ulich, Germany}
\affiliation{Institute of Physics, Johannes Gutenberg-University Mainz, 55128 Mainz, Germany}
\author{Stefan Bl\"ugel}
\affiliation{Peter Gr\"unberg Institut (PGI-1), Forschungszentrum J\"ulich and JARA, 52425 J\"ulich, Germany}   
\author{Phivos Mavropoulos}
\affiliation{Department of Physics, National and Kapodistrian University of Athens, Panepistimioupolis 15784 Athens, Greece}

\begin{abstract}
We present an \textit{ab initio} computational analysis of the topological Hall effect arising from stable magnetic skyrmions in the Pd/Fe/Ir(111) film using non-collinear spin density functional calculations within the Korringa-Kohn-Rostoker (KKR) Green function method. The semiclassical Boltzmann transport equation is employed for the resistivity and the Hall angle of the system. We explore the influence of the skyrmion size and the impact of disorder on the topological Hall angle.
\end{abstract}

\maketitle

\section{Introduction}
Magnetic skyrmions~\cite{skyrme_non-linear_1997,muhlbauer_skyrmion_2009, yu_real-space_2010, everschor-sitte_perspective_2018, bogdanov_physical_2020,gobel_beyond_2021,wang_fundamental_2022} have gained focus in many studies in recent years, as they are exceptionally promising for spintronics applications~\cite{fert_skyrmions_2013}. In ultrathin magnetic films, magnetic skyrmions are two-dimensional non-collinear magnetic spin textures that are topologically protected at the nanoscale~\cite{nagaosa_topological_2013}, i.e., they cannot be continuously deformed into another spin configuration without overcoming an energy barrier~\cite{rohart_path_2016}. Additionally, they are confined in space and can move as particles. Magnetic skyrmions can be formed at the interface of ferromagnetic films with heavy metals, as a consequence of the spin-orbit coupling (SOC) of conduction electrons and the broken space inversion symmetry, which promote the chiral symmetry breaking Dzyaloshinskii-Moriya interaction (DMI)~\cite{crepieux_dzyaloshinskymoriya_1998,heinze_spontaneous_2011}. Their special characteristics, as their creation, manipulation by electric currents, and detection, have been demonstrated both experimentally and theoretically, making them ideal candidates for computing applications~\cite{fert_skyrmions_2013,sampaio_nucleation_2013,kang_voltage_2016, yu_room-temperature_2017, fert_magnetic_2017, hoffmann_skyrmion-antiskyrmion_2021}.

Tunneling  and Hall effect methods offer promising prospects for the electrical detection of magnetic skyrmions. Among the former, typically applied in the current-perpendicular-to-plane geometry, in particular the non-collinear magnetoresistance effect~\cite{hanneken_electrical_2015,crum_perpendicular_2015,kubetzka_impact_2017,li_proposal_2024} takes advantage of the non-collinear skyrmion structure affecting the tunneling current between the sample and a scanning tunneling microscope tip. The latter method, which is applied in the current-in-plane geometry and refers to the measured response  transverse to an electric current and a perpendicularly applied magnetic field, is the subject of the present paper. In brief, Bloch electrons contributing to a current in a ferromagnet are skew-scattered by the chiral spin texture of magnetic skyrmions, resulting in a transverse contribution to the current that is detected as a Hall voltage. 

In more detail, the detection of skyrmions relies on the topological Hall effect~\cite{bruno_topological_2004,neubauer_topological_2009,kanazawa_discretized_2015,hamamoto_quantized_2015,gobel_unconventional_2017,maccariello_electrical_2018,denisov_topological_2018,Ishizuka:2018,Verma:2022} (THE) that  results from the non-coplanar magnetization  texture of skyrmions. In the limit of large skyrmions, one can adopt a semiclassical view of electron transport, where electrons move adiabatically through a magnetic skyrmion texture by which they gain a Berry phase proportional to the topological charge of the skyrmion, which contributes to an effective magnetic field that causes the transverse response of the electrons. The topological Hall effect is an additional contribution to the ordinary~\cite{Hall:1879} (OHE) and anomalous Hall effects~\cite{nagaosa_anomalous_2010} (AHE). The former is attributed to the Lorentz force resulting from an external magnetic field in nonmagnetic metals, while the latter occurs in collinear ferromagnets and results from the interplay between the intrinsic magnetization and spin-orbit coupling. Here the spin-orbit interaction causes an effective magnetic field in the reciprocal space leading to the transverse deflection of the electrons.

Deciphering the Hall transport properties in non-collinear textures remains a formidable challenge. For instance, an experimental separation between the anomalous and topological Hall effects is not trivial, as in chiral magnets and skyrmion hosting heterostructures a sizable spin-orbit interaction is required for the formation of skyrmions. A commonly practised  procedure to measure the topological Hall resistance makes use of the assumption  of a linear superposition of all Hall resistances, and thus  a linear superposition of spin-orbit and topological contributions. Employing this idea, one determines the topological Hall resistance as the difference of the  Hall resistances of the ferromagnetic sample with and without skyrmions~\cite{neubauer_topological_2009, Lee:2009}. The validity  of this heuristic assumption has recently been questioned in particular for systems with strong spin-orbit interaction such as heterostructures involving layers of $5d$-transition metals like Ir or Pt~\cite{Bouaziz:2021}.

In this work, we focus on the topological Hall effect that arises from stable magnetic skyrmions in a Pd/Fe bilayer on Ir(111). The fcc-Pd/Fe/Ir(111) thin film is a paradigmatic system for atomic-scale skyrmions in ultrathin films. Single N\'eel-type magnetic skyrmions with a size of about four nanometers in which the magnetisation textures winds clockwise from the inside to the outside (helicity $\pi$)  were  detected  in this system for the first time by spin-polarized scanning tunneling microscopy (STM) experiments at about 4~K and an external magnetic field of about 2~T~\cite{romming_writing_2013,romming_field-dependent_2015, leonov_properties_2016}. Additionally, this heterostructure has been extensively investigated using Density Functional Theory (DFT) simulations to assess its energetic stability~\cite{dupe_tailoring_2014, simon_formation_2014, crum_perpendicular_2015, lima_fernandes_universality_2018, fernandes_impurity-dependent_2020}. Today, we know that Pd can grow in an all fcc-stacking or in an hcp-stacking geometry leading to two different skrymions radii~\cite{kubetzka_impact_2017, von_malottki_enhanced_2017}. However, to the best of our knowledge, no \textit{ab initio} simulations have been conducted to study the Hall-type transport properties in this heterostructure. 

The conventional approach for the calculation of the topological Hall effect relies on the evaluation of the real-space Berry curvature~\cite{Franz:2014,Hamamoto:2015} acquired by the electrons following adiabatically a smoothly varying non-coplanar magnetization texture of the skyrmion.
Here, we take a different, challenging and hardly practised route and compute the electronic structure and the topological Hall effect  in skyrmion materials by considering entire nanosized skyrmions embedded in ferromagnets. This maybe an important route for small skyrmions subject to a large spin-orbit interaction or DMI, respectively. Our simulations are based on non-collinear spin-density functional theory (DFT)~\cite{U_von_Barth_1972} within the full-potential relativistic Korringa-Kohn-Rostoker Green function method (KKR)~\cite{papanikolaou_conceptual_2002,ebert_calculating_2011}. The DFT calculations are used to determine self-consistently the magnetic texture of the skyrmions, and allow the determination of the scattering rate of spin-polarized Bloch electrons at the skyrmion which then serves as an input to the Boltzmann formalism to investigate the topological Hall effect within a realistic description of the electronic structure imposed by the skyrmion in the fcc-Pd/Fe/Ir(111) system.

The Boltzmann formalism takes into account the elastic electron scattering by the skyrmionic topological magnetic structure. The effect of the non-coplanar magnetization direction is included in the scattering amplitude and scattering rate, although the Berry phase of the magnetization profile is not explicitly calculated. Also included is the scattering by the difference in the potential of the skyrmion with respect to the host (difference in the magnetization modulus, in the spin-orbit coupling, and in the non-magnetic part of the potential). Even though, by definition, the topological Hall effect is only caused by the non-coplanar magnetization profile, all the above scattering sources contribute to skew scattering and to the Hall angle, and cannot be disentangled in an obvious way. 

The above sources also contribute to conventional scattering, and thus to the resistivity. The skyrmion concentration in the Fe film enters only as a multiplicative factor to the scattering rate, because the skyrmions are experimentally known to be nucleated at random, thus the electron phase is lost during the multiple scattering among them. In addition, the growth of heterostructures inevitably leads to some degree of structural and chemical disorder. This is taken into account by an effective mean disorder parameter $\Gamma$.

Since the skyrmion is formally considered to be a defect, it gives an extrinsic contribution to the AHE. Our formalism focuses on scattering on the Fermi surface and does not include the ordinary, intrinsic AHE derived from the Berry phase of the band structure and the Fermi sea, even though the band structure calculations include the spin-orbit coupling.

The paper is organized as follows. In Section~\ref{Sec:formskyrmions} we present our DFT results on the pristine ferromagnetic film and on the formation of stable magnetic skyrmions in the Fe layer of this film. In Section~\ref{Sec:methodology}, we briefly describe the DFT formalism and the methodology for the spin-transport calculations combining the KKR  method with the Boltzmann transport equation. In Sec.~\ref{sec:discussion} we discuss the transport properties of the system, with emphasis on the dependence of the topological Hall effect on the skyrmion size and on additional disorder. Finally, we summarize our findings and main conclusions in Sec.~\ref{sec:conclusions}.

\section{Formation of stable magnetic skyrmions}\label{Sec:formskyrmions}
\begin{figure*}
\centering	
     \includegraphics[width=0.8\textwidth]{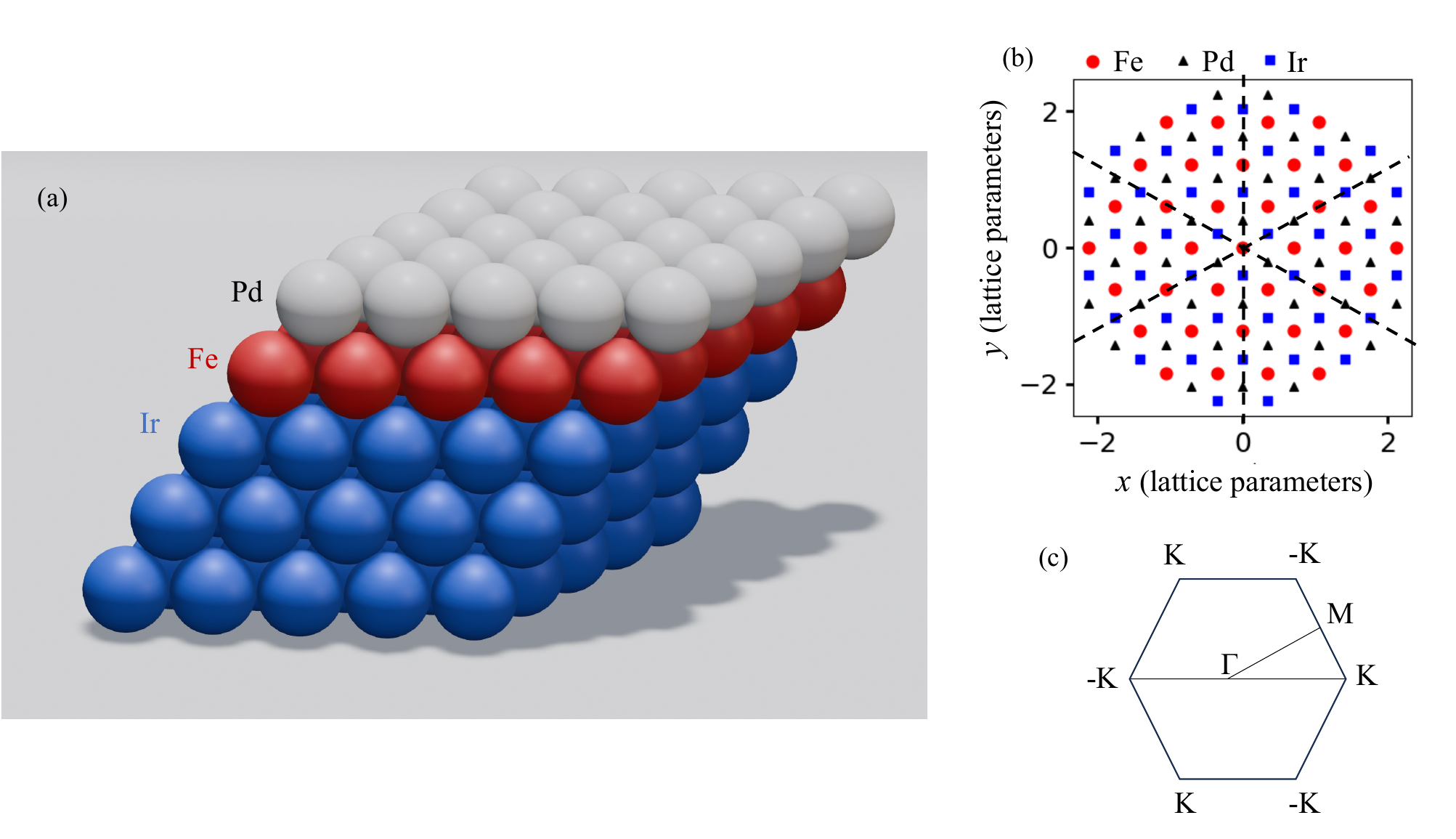}
	\centering	\includegraphics[width=0.8\textwidth]{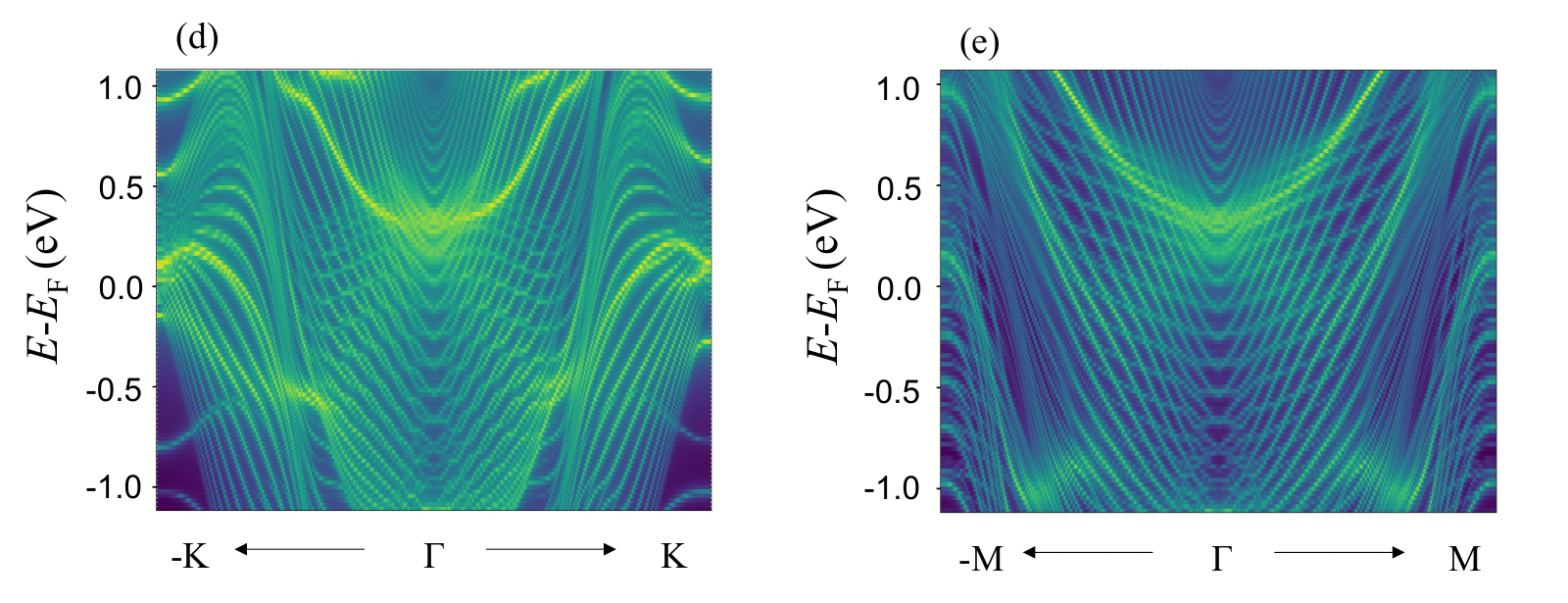}
	\caption{(a) Illustration of the fcc-stacked Pd/Fe/Ir(111) heterostructure. (b) Top-view illustration of the crystal structure of the near-surface atomic layers. The sites of Fe atoms and their nearest atoms, Pd above and Ir below, are represented by red circles, black triangles, and blue rectangles, respectively. The dashed lines represent the reflection planes perpendicular to the surface. The distance between two consecutive red dots (NN distance) is $(1/2)\sqrt{2}a$. The fcc lattice parameter and the NN distance are given in sec.\ IIIa. (c) The Brillouin zone of fcc(111), including the high-symmetry points $\Gamma$, K and M. (d) The band structure of a ferromagnetic 36-layer Pd/Fe/Ir(111) film for the spin-down-projected states in the high symmetry directions (d) $\mathrm{\Gamma}$K and (e) $\mathrm{\Gamma}$M. The line intensity corresponds to the degree of localization in the Fe layer.  }
	\label{fig:struc}
\end{figure*}

We start our analysis with the formation of stabilized magnetic skyrmions in the Pd/Fe/Ir(111) heterostructure based on \textit{ab initio} simulations. The system consists of a single atomic layer of ferromagnetic Fe deposited on a metallic Ir(111) substrate, and on top of this, there is an additional atomic layer of magnetically susceptible Pd metal, as depicted in Fig.~\ref{fig:struc}(a). The system is in an fcc(111) stacking geometry. In Fig.~\ref{fig:struc}(b), the crystal structure of the near-surface atomic layers is shown in a top view, i.e., the Pd layer, the Fe layer, and its nearest Ir layer. It is easily seen that there is 120\textdegree\ rotational symmetry and three reflection planes perpendicular to the surface. 

\begin{figure*}
	\centering	\includegraphics[width=0.9\textwidth]{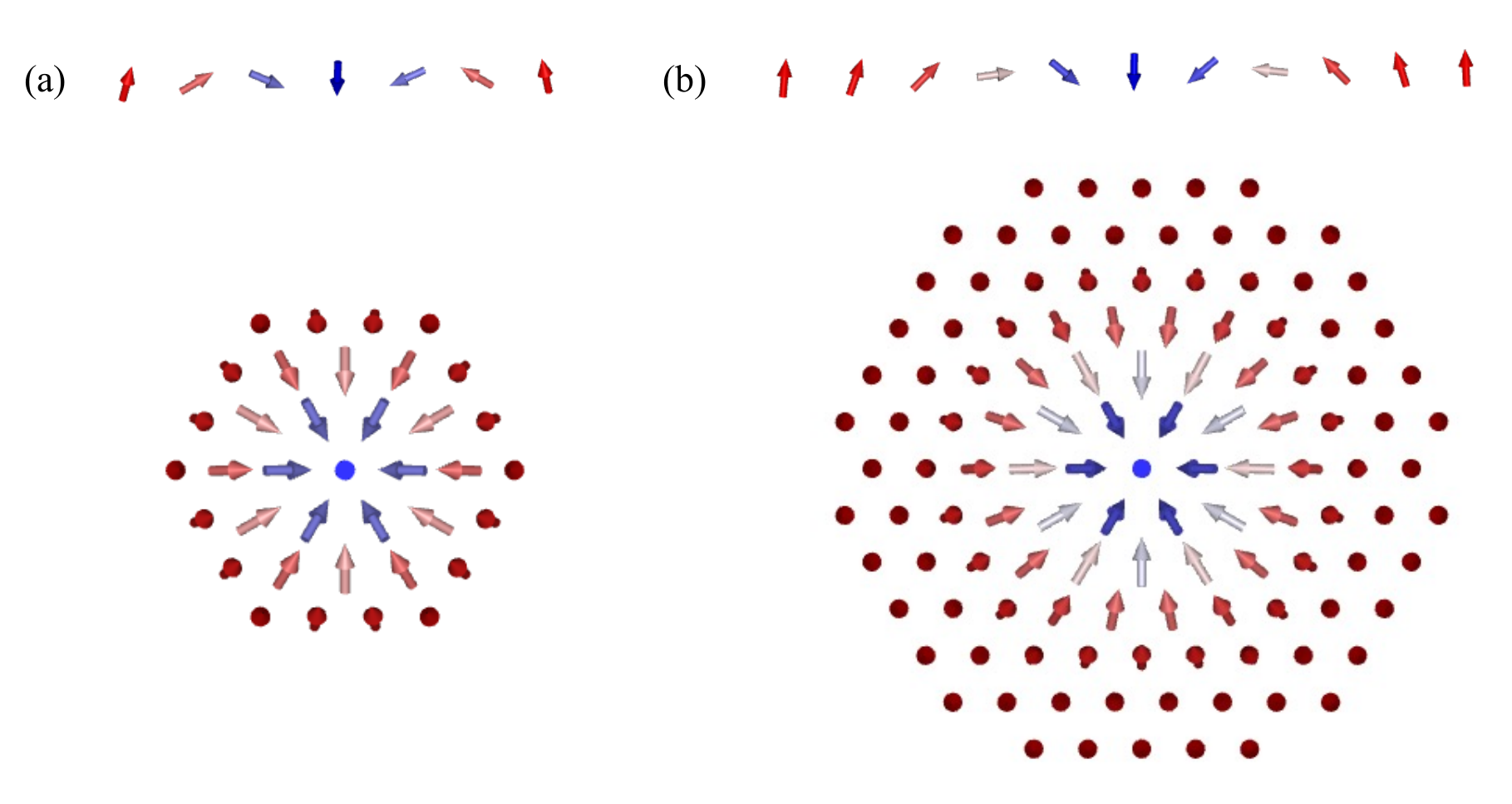}
	\caption{The relaxed magnetization at the Fe atoms within the two calculated skyrmions in the Pd/Fe/Ir(111) film. The skyrmions include (a) 37 Fe atoms and (b) 121 Fe atoms. The color code stands for the projection of the magnetization in $z$ direction.}
	\label{fig:skyrmform}
\end{figure*}

\begin{figure}
         \centering	\includegraphics[width=0.5\textwidth]{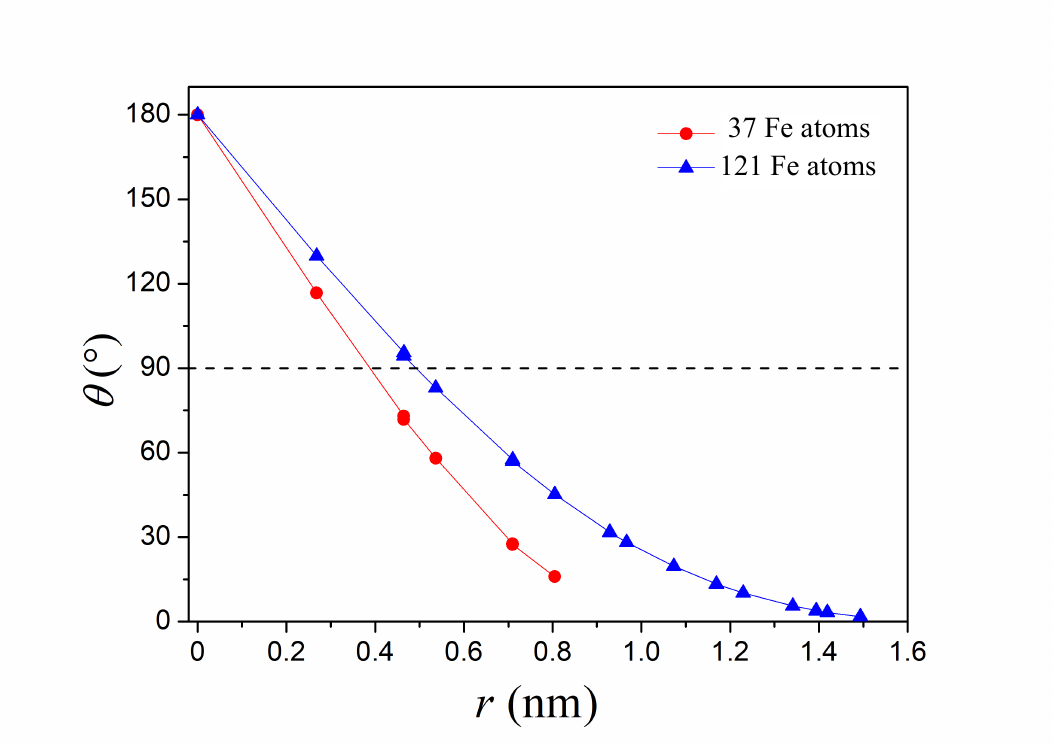}
	\caption{The profile of the polar angle of the magnetization, $\theta$, as a function of the distance, $r$, from the skyrmion center for the two calculated skyrmions. The red circles and blue triangles represent the skyrmions of 37 Fe and 121 Fe atoms, respectively. The distance at which $\theta=90$\textdegree\ is defined as the skyrmion radius, $r_{\mathrm{sk}}=0.385$ nm for the small skyrmion and $r_{\mathrm{sk}}= 0.5 $ nm for the large skyrmion.}
	\label{fig:skyrmion_radius}
\end{figure}

First, we show the electronic structure of the fcc-stacked ferromagnetic thin film Pd/Fe/Ir(111) of 36 layers thickness (34 Ir layers as structural model for the Ir substrate, 1 Fe and 1 Pd layer; structural details are given in Sec.\ \ref{Sec:methodology}). The magnetization is considered perpendicular to the plane. The high-symmetry points of the surface Brillouin zone are given in Fig.~\ref{fig:struc}(c). The band structure, projected to spin-down states within the Fe layer, is presented along the high-symmetry directions $\mathrm{\Gamma}$K and $\mathrm{\Gamma}$M in Figs.~\ref{fig:struc}(d),(e). Since spin-orbit coupling is present, all bands are of mixed spin character. However, the states that are of mainly spin-down character are displayed with a higher intensity. The same holds for the states that are more localized in the Fe layer.

In Fig.~\ref{fig:struc}(d) we observe a Rashba-type asymmetry, i.e., $E_{\bm{k}} \neq E_{\bm{-k}}$, in the $\Gamma-\mathrm{K}$ direction. The asymmetry is not very strong. It is well known and stems from the combined action of the exchange field, spin-orbit coupling and absence of space-inversion at the surface and promotes the Dzyaloshinskii-Moriya interaction, resulting in the formation of stable magnetic skyrmions in this system.

Secondly, we proceed with the calculations on the skyrmion formation. The skyrmion is formally treated as a single but large defect, embedded within a defined radius in the ferromagnetic system Pd/Fe/Ir(111). The defect consists of the same atoms as the host system, but with the magnetization deviating from the ferromagnetic state.  Self-consistent calculations are performed to achieve the relaxation of the magnetization in the non-collinear skyrmion state. Since we use DFT for the relaxation of the magnetization, the charge density also relaxes, albeit with only a small change with respect to the host charge density. This can lead to the all-electrical detection of skyrmions via the non-collinear magnetic resistance effect in an STM with a non-magnetic tip~\cite{hanneken_electrical_2015, crum_perpendicular_2015, kubetzka_impact_2017, li_proposal_2024}.

While in the experiment the skyrmion radius is controlled by the strength of an external magnetic field, in our calculations this is achieved by the use of the ferromagnetic boundary condition imposed at the Fe atoms of the host film in which the skyrmion is embedded. The skyrmion size is controlled by the number of atoms taken within the defect disk, i.e., by the region where the vector spin density (magnetization), length and direction of the atomic magnetic moments and charge density are allowed to relax, while being constrained by the boundary conditions. Obviously, one boundary condition is that the atomic potentials outside  the defect region are set in the ferromagnetic state with the magnetization perpendicular to the plane, i.e., the defect disk is considered as being perfectly embedded in a pristine ferromagnetic surface, without periodic boundary conditions. A second boundary condition is set for the central atom, which is spin flipped ($\theta=180$\textdegree) with respect to the ferromagnetic state. The skyrmion is established within the defect disc only, by the self-consistent first-principles relaxation of the magnetization in magnitude and direction, along with the convergence of the atomic potentials. This means that with a sufficiently large disk, the magnetization angle of the atoms at the edge of the disk does not appreciably deviate from the ferromagnetic state. A choice of a smaller disk moves the situation towards a smaller skyrmion, unrelaxed in size, reminiscent of the situation with a strong externally applied magnetic field that leads to smaller skyrmions~\cite{romming_field-dependent_2015}.

In our analysis, we examine two different skyrmion sizes. The smaller skyrmion is formed within a disk comprising 121 sites in the defect region. This includes the 1st to 5th in-plane neighbors of the central Fe atom, i.e., 37 Fe atoms in total. Additionally, the neighboring atoms of the topmost layer of the Ir substrate and of the Pd capping in the skyrmion region, i.e., 42 Ir and 42 Pd atoms are included in the self-consistent calculations. The larger skyrmion profile consists of 349 atoms in the defect region. It includes the 1st to 14th neighbors of the central Fe atom, resulting in 121 Fe atoms in total. This configuration also includes the nearest 114 Ir and 114 Pd atoms.

\begin{table}[htbp]
    \centering
 	\begin{tabular}{l c c c}
	    \hline
	    \mbox{}   & Ir & Fe & Pd \\  
		\hline\hline  
   $m_{s}$($\mu_{\mathrm{B}}$) & 0.04 & 2.53 & 0.29 \\
		\hline
		\hline
	\end{tabular}
	\caption{Calculated spin magnetic moment $m_{s}$ in the ferromagnetic state of Pd/Fe/Ir(111) for the topmost atomic layer of Ir, Fe, and Pd.}
	\label{magmomfmpdfeir}
\end{table}

In Figs.~\ref{fig:skyrmform}(a,b), we depict the relaxed spin structure for the two different skyrmion sizes. Obviously both skyrmions are found stable and both have relaxed to a clockwise helicity, in agreement with previous results from the literature~\cite{romming_writing_2013, dupe_tailoring_2014, crum_perpendicular_2015}. The corresponding profiles of the magnetization polar angle as function of the distance $r$ from the center of the skyrmion, $\theta(r)$, are presented in Figs.~\ref{fig:skyrmform}(a,b)(up) and Fig.~\ref{fig:skyrmion_radius}. The skyrmion radius is defined as the distance at which the magnetization angle reaches 90\textdegree. Thus, we estimate the skyrmion diameter of the smaller skyrmion equal to be $d_{\mathrm{sk}} = 0.77$ nm, and the larger $d_{\mathrm{sk}} = 1$ nm. The magnetization at the rim of the smaller skyrmion has relaxed at an angle of approximately $\theta$ = 16\textdegree, while of the larger skyrmion has relaxed at an angle $\theta$ = 1.8\textdegree, very close to the surrounding ferromagnetic state. Consequently, we have successfully achieved the relaxation of magnetization in the case of the larger skyrmion. 

The magnetic moments of Fe exchange proximitizes also a magnetization of Pd and Ir. We found that Fe imprints the clockwise helicity into the magnetization of Pd and Ir, i.e., the magnetic spin moments follow ferromagnetically polarized the magnetic spin moments of Fe. Since the exchange susceptibility of the Pd toplayer is larger than that of the Ir interface, the induced magnetic moments of Pd are much larger than those of Ir. The induced Ir moments are that small that we report here only on the average magnetic moment of the Ir layer interfacing Fe. The computed spin magnetic moment ($m_{s}$) of the near-surface layers of Ir, Fe, and Pd are presented in Table~\ref{magmomfmpdfeir}. We find that the calculated spin magnetic moment of the Fe atoms in the ferromagnetic background is large, while the induced magnetic moment of Pd is also non-negligible. Within the skyrmion region, the modulus of the spin magnetic moments of the Fe atoms ranges from $2.47~\mu_{\mathrm{B}}$ to $2.52~\mu_{\mathrm{B}}$ for the small skyrmion of diameter $d_{\mathrm{sk}} = 0.77$ nm, and from $2.49~\mu_{\mathrm{B}}$ to $2.53~\mu_{\mathrm{B}}$ for the large skyrmion of diameter $d_{\mathrm{sk}} = 1$ nm. Concerning Pd,  the spin magnetic moment modulus ranges from $0.15~\mu_{\mathrm{B}}$ to $0.27~\mu_{\mathrm{B}}$ for the small skyrmion and from $0.19~\mu_{\mathrm{B}}$ to $0.28~\mu_{\mathrm{B}}$ for the large skyrmion.

\section{Methodology}\label{Sec:methodology}

\subsection{Computational details}

The host system Pd/Fe/Ir(111) is modelled as a thin film consisting of 36 atomic layers in total, with 34 layers of Ir, 1 layer of Fe, and 1 layer of Pd in the fcc(111) stacking geometry. The configuration is completed by 3 vacuum layers on each side of the film. The fcc lattice constant used for the calculations is $a=3.793$~\AA,  corresponding to an in-plane nearest neighbor (NN) distance of 2.682~\AA. The interlayer distances were $d_{\mathrm{Ir-Ir}}$=2.19~\AA, $d_{\mathrm{Fe-Ir}}$=2.06~\AA, $d_{\mathrm{Pd-Fe}}$=2.02~\AA. All lattice parameters were taken from the work of Dup\`{e} et al.~\cite{dupe_tailoring_2014}, who \textit{ab initio} structurally optimised the parameters using lattice relaxations.

We employ density functional theory (DFT) within the local density approximation (LDA)~\cite{kohn_self-consistent_1965,vosko_accurate_1980}, utilizing  the full-potential relativistic Korringa-Kohn-Rostoker (KKR) Green function method~\cite{papanikolaou_conceptual_2002,ebert_calculating_2011}
as implemented in the J\"ulich full potential relativistic KKR code~\cite{jukkr2022,rusmann_aiida-kkr_2021}. The self-consistent solution of the Green function includes the spin-orbit coupling as well as scalar relativistic corrections. For the computation of the Green functions a finite angular momentum cutoff of $ l_{\mathrm{max}}=2 $ was used, which results in matrices of rank $N_{\rm atom}\times2(l_{\rm max}+1)^2=6282$ for the solution of the Dyson equation at each point in the energy integration contour (here, $N_{\rm atom}=349$ is the number of sites in the defect cluster). For the self-consistent calculations 35 energy points were used. The non-collinear DFT calculations for the self-consistent potential of the skyrmion atoms were carried out with the Jülich KKR impurity-embedding code KKRimp~\cite{bauer_development_2014}.

The Fermi surface, the group velocity at the Fermi surface, the Boltzmann formalism and the spin-transport  calculations were carried out using the PKKprime code~\cite{jukkr2022, zimmermann_ab_2014, zimmermann_fermi_2016}. We use 41700 $ k $-points on the full two-dimensional Fermi surface (Fermi line) of the Pd/Fe/Ir film. Here, for computational time reasons, the film thickness is restricted to 23 layers in total, i.e., 17 Ir atomic layers, 1 Fe layer, 1 Pd layer and 2 vacuum layers at each surface.

\subsection{Self-consistent calculations of skyrmions}

The charge and spin density within the skyrmion region are found in terms of the space-diagonal part of the Green function, $G^{\mathrm{Sk}}(\bm{r},\bm{r}';E)$ at $\bm{r}=\bm{r}'$, which involves the solution of  the 
Dyson equation:
\begin{align}
    G^{\mathrm{Sk}}&(\bm{r},\bm{r}';E) =  G^{\mathrm{host}}(\bm{r},\bm{r}';E) + \nonumber\\ & \int G^{\mathrm{host}}(\bm{r},\bm{r}'';E)\Delta V(\bm{r''})  G^{\mathrm{Sk}}(\bm{r}'',\bm{r}';E)\,d^3r''.
\end{align}
This equation relates the Green function of the ferromagnetic host system, $G^{\mathrm{host}}$, to the Green function of the perturbed non-collinear system, $G^{\mathrm{Sk}}$, through a perturbing potential $\Delta V$. Since the potential difference $\Delta V$, as well as the difference in charge and spin density, are confined in the skyrmion region, the integration in the Dyson equation and the sought-for Green function $G^{\mathrm{Sk}}(\bm{r},\bm{r}';E)$ are also confined in this region. Following the KKR formalism, the Dyson equation is transformed into a linear inhomogeneous system of algebraic equations and solved for $E$ on a contour in the complex energy plane. $G^{\mathrm{Sk}}$, $G^{\mathrm{host}}$, and $\Delta V$ are matrices in spin space, since we face a non-collinear magnetic problem. 

$\Delta V$ is the difference of the potential in response to the rotation of the magnetization. Consequently, it is related to the difference between the magnetization of the skyrmion and the magnetization in the ferromagnetic state. In a first approximation, it reads in atom $i$
\begin{equation}
    \Delta V_{i}(\bm{r}) \approx B_{i}(\bm{r}) \,(\hat{\bm{e}}_{\bm{M}_{i}}\vdot\bm{\sigma}  -  \hat{\bm{e}}_{z} \vdot\bm{\sigma} ),
\end{equation}
with $B_{i}$ the difference of the spin up and spin down components of the potential in the ferromagnetic state, i.e., $B_{i}(\bm{r})=V_{i}^{\uparrow}(\bm{r})-V_{i}^{\downarrow}(\bm{r})$, and $\bm{\sigma}=(\sigma_x,\sigma_y,\sigma_z)$ the Pauli vector, where spin-up and spin-down are taken with respect to the local spin axis. However, after full relaxation, the potential entails a change with respect to the ferromagnetic state also in $B_{i}(\bm{r})$ and in the scalar potential $\frac{1}{2}[V_{i}^{\uparrow}(\bm{r})+V_{i}^{\downarrow}(\bm{r})]$. The directions $\hat{\bm{e}}_{\bm{M}_{i}}$ are found self-consistently, but respect the boundary condition of being opposite to the direction of the ferromagnetic magnetization ($\theta=180$\textdegree) for the atom in the center and parallel to the ferromagnetic magnetization ($\theta=0$) for the atoms outside the skyrmion rim.  In the non-collinear magnetic skyrmion state, we adopted the approximation that the spin part of the potential, $B_i(r)\,\hat{\bm{e}}_{\bm{M}_{i}}\vdot \bm{\sigma}$, is collinear within the atomic cell $i$, but the direction $\hat{\bm{e}}_{\bm{M}_{i}}$ varies between different atomic cells.

\subsection{Topological Hall effect calculations}\label{sub:boltzmann}

Following previous work~\cite{mertig_transport_1999,gradhand_extrinsic_2010,long_spin_2014,kosma_strong_2020}, we proceed with the Boltzmann transport equation approach.

The transition matrix $T$ describing the scattering of a given Bloch state $\psi_{\bm{k}'}$ in the unperturbed ferromagnetic host heterostructure to a final state $\psi_{\bm{k}}$ is given by the matrix elements
\begin{equation}\label{tmatrix}
T_{\bm{k}\bm{k}'} = \int   \psi^{\dagger}_{\bm{k}}(\bm{r}) \Delta V(\bm{r}) \psi^{\mathrm{Sk}}_{\bm{k}'}(\bm{r})\,d^3r,
\end{equation}
where the state scattered by the skyrmion obeys the Lippmann-Schwinger equation
\begin{eqnarray}
\psi^{\mathrm{Sk}}_{\bm{k}'}(\bm{r}) \! &=&\! \psi_{\bm{k}'}(\bm{r}) + \!\int G^{\mathrm{host}}(\bm{r},\bm{r}';E_{\bm{k}'}) \Delta V(\bm{r}') \psi^{\mathrm{Sk}}_{\bm{k}'}(\bm{r}') d^3r' \nonumber\\
\! &=&\! \psi_{\bm{k}'}(\bm{r}) + \!\int G^{\mathrm{Sk}}(\bm{r},\bm{r}';E_{\bm{k}'}) \Delta V(\bm{r}') \psi_{\bm{k}'}(\bm{r}') d^3r',\nonumber\\
& &
\end{eqnarray}
and where the region of integration is restricted to the skyrmion defect. Since the full Green function enters in the expression for $\psi^{\mathrm{Sk}}_{\bm{k}'}$, Eq.\ (\ref{tmatrix}) expresses the full $T$-matrix, which is equivalent to summing up all multiple scattering events among the skyrmion atoms to all orders of perturbation theory.
For the scattering rate off a single skyrmion, $w_{\bm{kk'}}$, we employ the Golden Rule
\begin{equation}\label{fermigoldenrule}
w_{\bm{kk'}} = \frac{2\pi}{\hbar} \delta(E_{\bm{k}} - E_{\bm{k}'}) \abs{T_{\bm{k}\bm{k'}}}^{2}.
\end{equation}

In a next step, the semi-classical linearized Boltzmann transport equation is solved self-consistently. This provides the vector mean free path $\mathbf{\Lambda}_{\bm{k}}$, on the Fermi surface: 
\begin{equation}\label{selfconspath}
\mathbf{\Lambda}_{\bm{k}} \vdot \hat{n}_{\bm{\mathcal{E}}} = \ \tau_{\bm{k}} \bigg[\bm{v_{\bm{k}}} \vdot \hat{n}_{\bm{\mathcal{E}}}+ \sum_{\bm{k'}} w_{\bm{k}\bm{k'}} (\mathbf{\Lambda}_{\bm{k'}} \vdot \hat{n}_{\bm{\mathcal{E}}})\bigg].
\end{equation}
Here, $ \bm{v_{\bm{k}}} $ represents the group velocity, $ \hat{n}_{\bm{\mathcal{E}}} =  \bm{\mathcal{E}}/\abs{\bm{\mathcal{E}}}$ is the direction of the electric field $ \bm{\mathcal{E}} $, and $\tau_{\bm{k}}$ is the relaxation time, defined as 
\begin{equation}
    \tau_{\bm{k}} = 1/ \sum_{\bm{k'}} w_{\bm{k'}\bm{k}}.
    \label{eq:relaxtime}
\end{equation}
The Boltzmann equation is self-consistently solved beyond the relaxation time approximation, and it includes  the ``scattering-in term,'' accounting for the vertex corrections~\cite{butler_theory_1985,gradhand_extrinsic_2010}.

The focus of this work is on the investigation of the topological Hall effect. In order to calculate the Hall angle and analyze the topological Hall effect, it is essential to compute the conductivity tensor, which is expressed in terms of the vector mean free path via 
\begin{equation}\label{conductivitytensor}
\sigma_{ij} = \frac{e^{2}}{4 \pi^{2}} \int_{\mathrm{FS}} \frac{dk}{\hbar \abs{\bm{v_{\bm{k}}}}} (\bm{v_{\bm{k}}})_{i}\ ( \mathbf{\Lambda}_{\bm{k}})_{j},
\end{equation}
with $i$ and $j$ $\in\{x, y\}$. As a result, we are able to calculate the resistivity tensor $\bm{\rho} = \bm{\sigma}^{-1}$. We can now also determine the Hall angle, denoted as $\alpha = \sigma_{xy}/\sigma_{yy}$, or, in terms of the resistivity, $\alpha \approx \rho_{yx}/\rho_{xx}$ when considering  $\sigma_{xy}=-\sigma_{yx}$ and assuming to a good approximation $\sigma_{xx}=\sigma_{yy}$ because the  highly isotropic nature of the $C_{3v}$ point symmetry of the system,  assuming that the electric  field is in the $y$ direction perpendicular to the surface normal in the Pd/Fe/Ir heterostructure. Within this approach, the scattering among different skyrmions is only taken into account on the average by introducing a skyrmion concentration parameter $c_\textrm{Sk}$ that multiplies the scattering rate $w_{\bm{k}\bm{k'}} \rightarrow c_\textrm{Sk} w_{\bm{k}\bm{k'}}$. The inverse mean free path and the inverse conductivity tensor scale linearly with the concentration, thus the concentration cancels in the expression for $\alpha$.

\begin{figure*}
    \centering
    \includegraphics[width=0.9\textwidth]{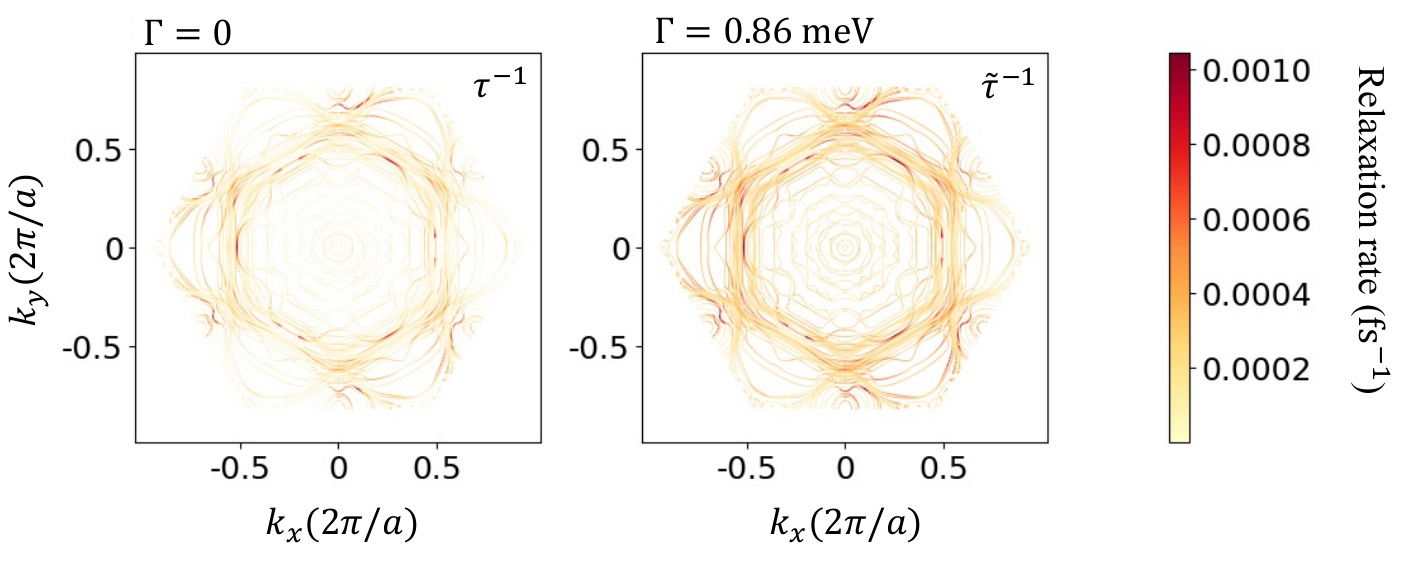}
    \caption{The relaxation rate $\tau_{\bm{k}}^{-1}$ (Eq.\ \ref{eq:relaxtime}) for the states at the Fermi surface, for the full case of the large skyrmion (case iii). The left panel represents the case without disorder, while the right panel depicts the case with a disorder strength of $\Gamma=0.86$~meV.}
    \label{fig:scattering_rate}
\end{figure*}

\begin{figure}[htbp]
    \centering
    \includegraphics[width=0.5\textwidth]{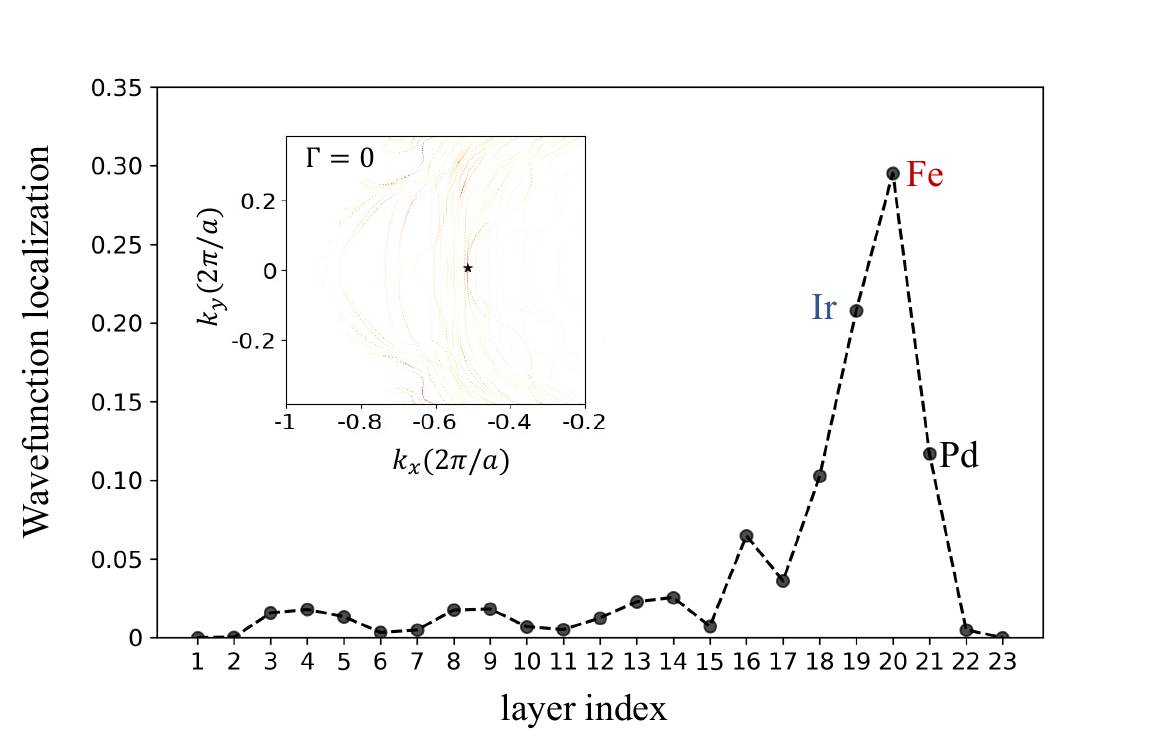}
    \caption{The wavefunction localization in the atomic layers of the Pd/Fe/Ir(111) film, for a $\bm{k}$ point in a region with high scattering intensity [$\bm{k}=(-0.514, 0)\,2\pi/a$]. The inset zooms into a region of Fig.~\ref{fig:scattering_rate}(a), highlighting the k-point, which is marked with a star.}
    \label{fig:dos_taumax}
\end{figure}

\subsection{\label{disorder}Disorder induced scattering}

In a realistic situation, there are scattering sources besides the skyrmions, e.g., impurity atoms, structural defects, phonons, thermal fluctuations  etc., that are not precisely known but affect the transport properties. The latter two scattering sources are not so important for this study as the experiments are carried out at 4~K. It is important to consider this residual scattering, since otherwise the substrate will show no resistance, causing a short-circuit. 

A simple way to account for such disorder is by introducing an ad-hoc scattering rate~\cite{geranton_spin-orbit_2016}. The relaxation time $\tau_{\bm{k}}$ (Eq.~\eqref{eq:relaxtime}) is adjusted to $\Tilde{\tau}_{\bm{k}}$ with the following relation:
\begin{equation}\label{taudis}
    \frac{1}{\Tilde{\tau}_{\bm{k}}} = c_\textrm{Sk}\frac{1}{\tau_{\bm{k}}} + \frac{2\Gamma}{\hbar}.
\end{equation}
Here, the parameter $\Gamma$ denotes the disorder strength with energy dimensions of the order of meV. Consequently, the matrix elements of the scattering rate $w_{\bm{k'k}}$ are modified to ensure the consistency of $\Tilde{\tau}_{\bm{k}}=1/\sum_{k'}\Tilde{w}_{\bm{kk'}}$. The adjusted scattering rate $\Tilde{w}_{\bm{k'k}}$ is determined using the equation~\cite{geranton_spin-orbit_2016}:
\begin{equation}\label{pkkdis}
    \Tilde{w}_{\bm{kk'}} =  c_\textrm{Sk} w_{\bm{kk'}}+\frac{2\Gamma}{\hbar n(E_{\bm{k}})} \delta(E_{\bm{k}}-E_{\bm{k}'}),
\end{equation}
where $n(E)$ is the density of states.

Subsequently, the Boltzmann equation (Eq.~\eqref{selfconspath}) is solved, taking into account the constant energy term via the modified relaxation time $\Tilde{\tau}_{\bm{k}}$ and the adjusted scattering rate $\Tilde{w}_{\bm{kk'}}$.

\section{Results and Discussion}\label{sec:discussion}

In the transport calculations, we consider three different cases as numerical experiments: 
\begin{enumerate}[i]
\item the case of a small skyrmion of 37 Fe atoms, 
\item the case of a large skyrmion of 121 Fe atoms, but taking into account the electron scattering only by the perturbed potential of the Fe atoms, and 
\item the full case of a large skyrmion of 121 Fe atoms, taking into account the electron scattering also by the perturbed potential of the Ir substrate and Pd capping in the skyrmion region. This case actually corresponds to our best approximation to a fully relaxed skyrmion magnetization. 
\end{enumerate}

As a numerical experiment, our calculations include five different skyrmion concentrations (0.1$\%$, 2$\%$, 4$\%$, 6$\%$, and 10$\%$) in the Pd/Fe/Ir surface. It should be noted that, among these concentrations, the only realistically attainable one is the lowest, 0.1$\%$.  Since a skyrmion encompasses a number of the order of 100 Fe atoms (121 atoms, in our calculation of the skyrmion with 1 nm diameter), a 1\% concentration already corresponds to full coverage, as one has e.g.\ in the case of skyrmion lattice in the Fe/Ir(111) system~\cite{heinze_spontaneous_2011}. Consequently, higher concentrations serve as a check for the consistency of our calculations.

\subsection{Relaxation rate}\label{ssec:RR}
We start with a discussion of the scattering properties.
Fig.\ \ref{fig:scattering_rate} shows the relaxation rate $1/\tau_{\bm{k}}=\sum_{\bm{k'}}w_{\bm{k'}\bm{k}}$ (see Eq.\ \eqref{eq:relaxtime}) in a color scale for the states at $E_{\bm{k}}=E_{\mathrm{F}}$, for the full case (iii) of the large skyrmion.
In the left panel, the relaxation rate is shown for the case without disorder, i.e. $\Gamma=0$, while, in the right panel, $\Gamma=0.86$\ meV is taken, resulting in an obvious increase of the relaxation rate overall. We see that certain parts of the Fermi surface are rather pronounced in scattering intensity (red-shifted regions). These regions correspond to states that are strongly scattered by the skyrmions, hence they must have a high degree of localization near the surface. This hypothesis is tested in Fig.~\ref{fig:dos_taumax}, which depicts the wavefunction localization $|\psi_{\bm{k}}(\bm{r})|^2$ in each atomic layer in the Pd/Fe/Ir(111) system, for a chosen $\bm{k}$ point within a region of high scattering intensity (shown in the inset of Fig.~\ref{fig:dos_taumax}). The peak in the Fe layer, and the high degree of localization in the Pd and Ir surface layers, reveal a surface resonance.

\subsection{Longitudinal resistivity}\label{ssec:LR}
\begin{figure}[htbp]
    \centering
    \includegraphics[width=0.5\textwidth]{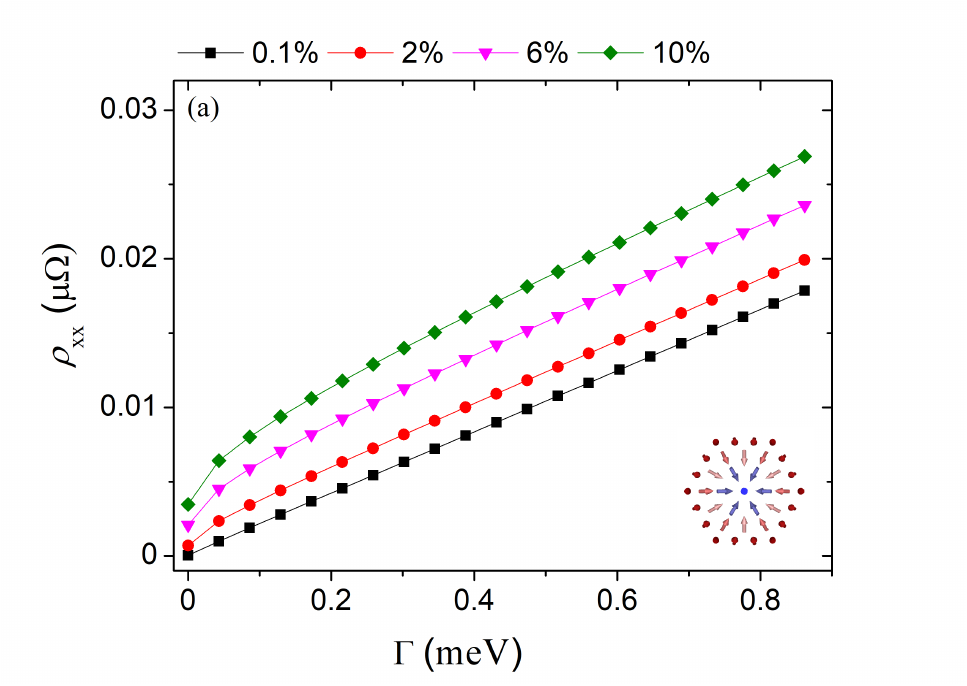}
    \includegraphics[width=0.5\textwidth]{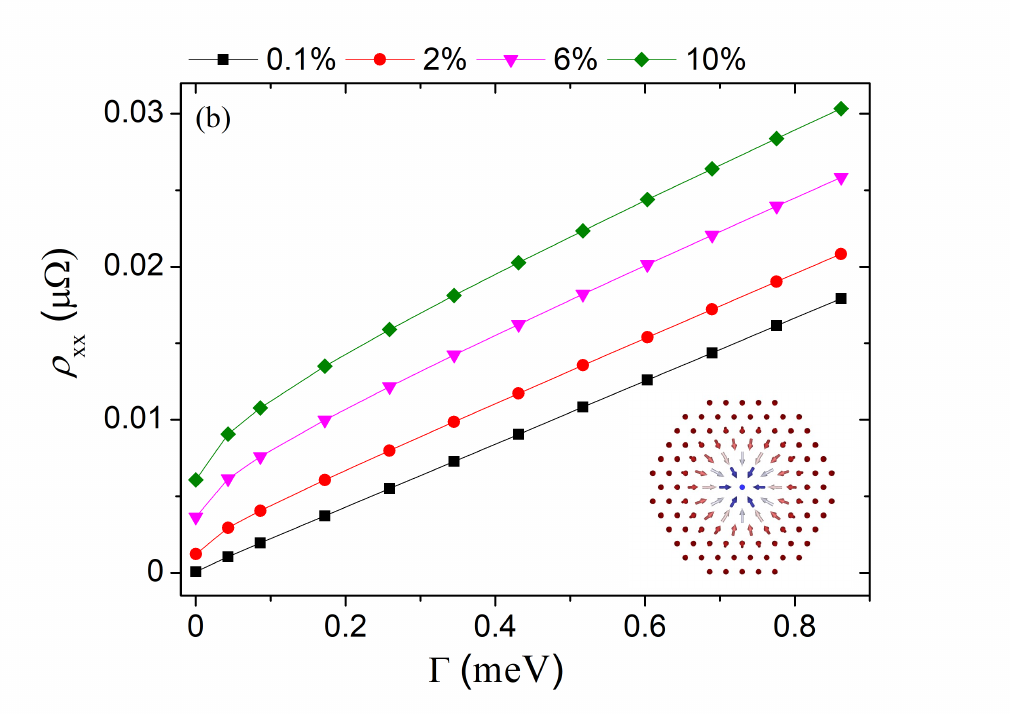}
    \includegraphics[width=0.5\textwidth]{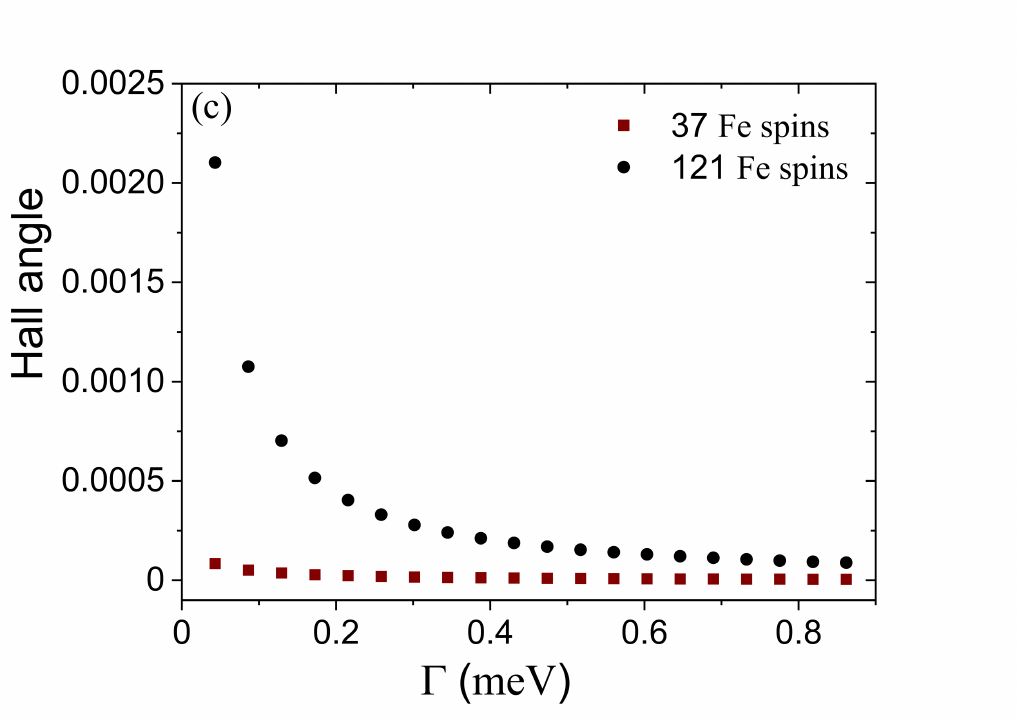}
    \caption{The longitudinal resistivity $\rho_{xx}$ as a function of the disorder parameter $\Gamma$ for the skyrmion of diameter (a) $0.77$ nm (case i) and (b) $0.99$ nm (case ii) in Pd/Fe/Ir(111) film. The black squares, red circles, pink down-triangles and green rhombus correspond to four distinct skyrmion concentrations, 0.1$\%$, 2$\%$, 6$\%$ and 10$\%$ in the Pd/Fe/Ir(111) film, respectively. (c) The topological Hall angle as a function of $\Gamma$ for the skyrmions of diameter $0.77$ nm (case i) and $1$ nm (case ii) in Pd/Fe/Ir(111) film, considering a 0.1$\%$ skyrmion concentration in the system surface.}
    \label{fig:rho_37121_fespins}
\end{figure}

We now examine the impact of the disorder strength $\Gamma$ (Eq.~\eqref{taudis}) on the longitudinal resistivity $\rho_{xx}$. The results of our calculations are shown in Figs.~\ref{fig:rho_37121_fespins}(a,b) for the smaller and larger skyrmion system considering the electron scattering only by the Fe atoms within the skyrmion, i.e., cases i and ii, respectively. 
Since we have two sources of scattering, the skyrmions and the disorder $\Gamma$, we expect that some resistivity will survive at $\Gamma$$=$$0$, which henceforth we call residual resistivity.

We observe that in the case of a realistic scenario with $0.1\%$ skyrmion concentration in the surface, the longitudinal resistivity $\rho_{xx}$ exhibits a linear dependence on the disorder. Here, the skyrmion induced scattering is very low, so that the skyrmion contribution to the residual resistivity is below point thickness in the plot. However, at higher skyrmion concentrations, the skyrmion contribution to the residual resistivity is apparent. At these concentrations we find that the resistivity as a function of $\Gamma$ changes slope as $\Gamma$ increases from zero, but the slope is then stabilized, at higher $\Gamma$, as disorder becomes the main source of scattering.  

Comparing the two different skyrmion sizes, we see that the residual resistivity is different, but is of the same order of magnitude. This is expected, as the larger skyrmion produces more scattering.

\subsection{Hall angle}\label{ssec:HA}
Next, we discuss our results on the Hall angle. The calculated values comprise the contributions of (i) the extrinsic anomalous Hall effect as it arises from impurity skew-scattering in a ferromagnetic host, (ii) the spin-orbit interaction of the non-collinear skyrmion texture in difference  to the ferromagnetic state, which gives rise to the anomalous Hall effect, (iii) as well as the topological Hall effect, as expected from the fact that the impurity here (full skyrmion) has a non-coplanar  spin texture. The formalism includes all contributions (i-iii) and, similar to the experiment, cannot distinguish between them. However, it is clear that the main source of scattering by a skyrmion is its spin texture, while the scattering due to the difference in charge density or spin-orbit coupling compared to the pristine ferromagnetic surface will be comparably insignificant. Therefore, we consider the calculated Hall effect to be a topological Hall effect, at least in the limit of low disorder ($\Gamma\rightarrow0$) and low spin-orbit interaction.

The behavior of the Hall angle $\alpha$ as a function of $\Gamma$ is illustrated in Fig.~\ref{fig:rho_37121_fespins}(c) for the two different-sized skyrmion systems i and ii. Here, we present the results only for the realistic skyrmion concentration of 0.1$\%$. While the longitudinal resistivity is of the same order of magnitude for the two skyrmion sizes, a significant increase in the Hall angle in the larger skyrmion system is observed at low $\Gamma$. The latter suggests a significant contribution from the off-diagonal conductivity term.  We also see that the Hall angle becomes relatively small as the disorder strength increases. 

\begin{figure}[htbp]
    \centering
    \includegraphics[width=0.5\textwidth]{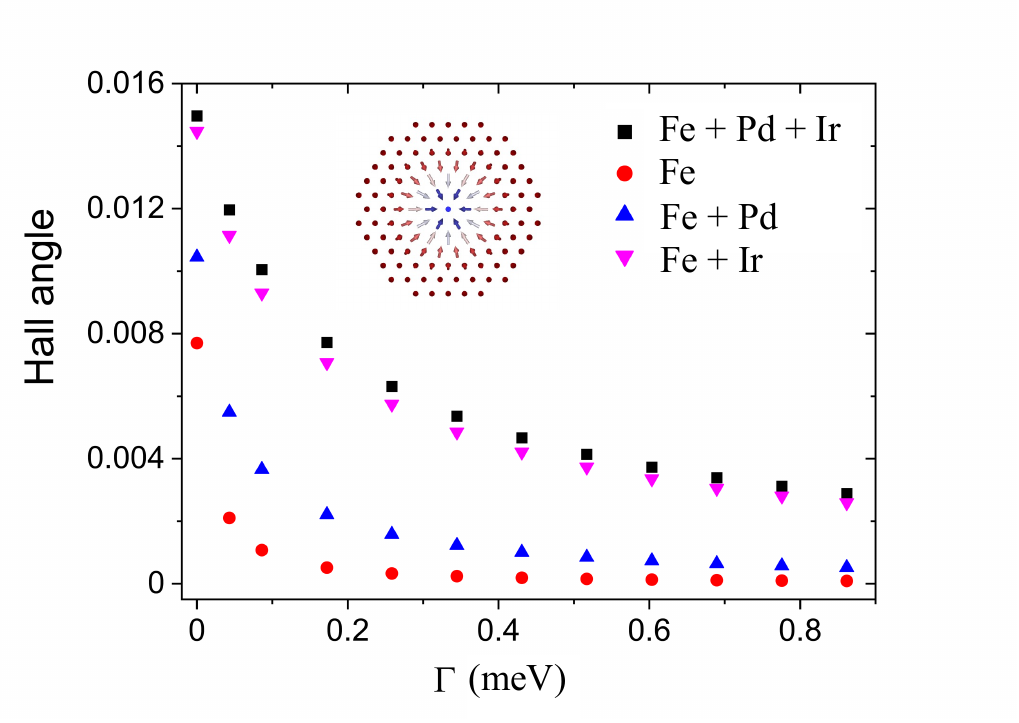}
    \caption{The topological Hall angle as a function of the disorder parameter $\Gamma$ for the skyrmion of a diameter of $1$ nm and a concentration of 0.1$\%$ in the Pd/Fe/Ir(111) surface. The black squares correspond to calculations that take into account the perturbed potential of the Ir substrate and Pd capping in the skyrmion region (case iii). The red circles represent the case where only the perturbed potential of the Fe atoms is considered (as in Fig.~\ref{fig:rho_37121_fespins}(c)). The blue up-triangles correspond to the case where only the perturbed potential of the Pd capping and Fe atoms is considered, and the pink down-triangles represent the case where only the perturbed potential of the Ir substrate and Fe atoms are included.}
    \label{fig:rhotha_349}
\end{figure}

We now discuss case (iii), i.e., the large skyrmion including the contributions of the neighboring atoms of the topmost Ir layer and of the Pd capping layer in the scattering calculations. The results are  shown in Fig.~\ref{fig:rhotha_349} (black squares). We observe a significant increase in the calculated Hall angle, as compared to the corresponding results obtained when only the Fe atoms in the skyrmion region are considered during the scattering process (Fig.~\ref{fig:rhotha_349} (red circles)). We also examine the cases where the electron scattering is considered only by the perturbed potential of the topmost Ir and Fe atoms (Fig.~\ref{fig:rhotha_349} (pink down-triangles)), or only by the perturbed potential of the Pd and Fe atoms. We find that the contribution of Ir is more significant than that of Pd to the topological Hall angle, likely due to the strong spin-orbit coupling of the Ir substrate. We conclude that the layers with high spin susceptibility, Ir and Pd, are polarized by Fe, contribute to the spin texture, and thus play a significant role in the Hall effect.

In this improved model, the decrease of the Hall angle as a function of the disorder parameter is also seen. This result underscores the significant impact of the degree of disorder in the sample on the measured Hall angle. Therefore, the inclusion of impurity atoms and other scattering sources is crucial for accurately describing the topological Hall effect.

\section{Summary and Conclusions}\label{sec:conclusions}
In this work we have taken up the challenge of using \textit{ab initio} theory to shed light onto the understanding of the Hall effect of non-collinear magnetic textures. For this purpose we have chosen the experimentally and theoretically well-studied fcc-Pd/Fe bilayer on Ir(111). It is a system with strong spin-orbit interaction caused by the $5d$-metal Ir and exhibits small, magnetic nanoscale skyrmions in the Fe film.  We deal with the Hall effect by considering the skew scattering of Bloch electrons from a skyrmion perfectly embedded as a single defect in the ferromagnetic film, determining the magnetic and electronic structure of the whole skyrmion self-consistently, and relating the scattering results in the form of scattering rates to mean free paths and the conductivity tensor employing the Boltzmann equation. The scattering from single atomic defects or other perturbations, which lead to the typical resistances in the longitudinal transport, were phenomenologically taken into account by a constant parameter $\Gamma$. Our main focus is on the topological Hall effect (THE), which we discuss in terms of the Hall angle. Although the THE is the main contribution to the Hall effect in skyrmions, in our calculation, similar to the experiment, the Hall effect due to topology and spin-orbit interaction are interwound.

In summary, we applied the full-potential relativistic KKR Green function method within non-collinear spin density functional  calculations to the formation of stable magnetic skyrmions in the Fe ferromagnetic layer of the fcc-Pd/Fe/Ir(111) heterostructure. We modelled the skyrmion  by including 37 Fe atoms and 121 Fe atoms self-consistently in the scattering regions. In both cases the skyrmion radius was found to be about 0.5~nm.  Subsequently, a calculation of the electron scattering rate off the skyrmions was used as input to the semiclassical Boltzmann equation for the calculation of the resistivity and the topological Hall angle. 

Delving into the electron scattering details of the system, we conclude that surface states and surface resonances scatter strongly off the (surface-localized) skyrmions, and are therefore most significant for the transport properties.

We find four important factors that significantly affect the results and therefore should be accounted for in realistic simulations. First, judging from the two different skyrmion sizes that we analyzed, we infer that the skyrmion size plays an important role. Second, one should incorporate in the transport calculations the non-collinear magnetization that is induced to the Ir and Pd layers by the skyrmion state in the Fe layer. Both significantly increase the robustness of the topological Hall angle against disorder. The large spin-orbit interaction of Ir combined with the small spin-polarization (0.04~$\mu_\textrm{B}$) is a more important factor than the large spin-polarization of Pd (0.29~$\mu_\textrm{B}$). Third, the large contribution of the spin-orbit interaction to the topological Hall angle raises the question of disentangling the two contributions. This supports the question of whether the disentanglement leads to additional contributions to the Hall effect beyond the AHE and THE, such as the chiral~\cite{Lux:2020} and non-collinear Hall effect~\cite{Bouaziz:2021}. Fourth, disorder suppresses the Hall angle. Thus, the degree of disorder from different sources (e.g.\ impurity atoms) must not be neglected. 
\
\section{Acknowledgements}
We would like to thank Juba Bouaziz for fruitful discussions. The research work was supported by the Hellenic Foundation for Research and Innovation (HFRI) under the HFRI PhD Fellowship grant (Fellowship Number: 1314). We acknowledge support by the German Academic Exchange Service (DAAD) and the EC Research Innovation Action under the H2020 Programme under the Project HPC-EUROPA3 (HPC17CTZOF). A.K, P.R. and S.B. acknowledge funding from the Deutsche Forschungsgemeinschaft (DFG) through ML4Q (EXC 2004/1 – 390534769). This work was supported by computational time granted from the Greek Research \& Technology Network (GRNET) in the National HPC facility-ARIS-under project ID ``pr007039-TopMagX,'' and the HLRS. 

\bibliography{ref}
\end{document}